\documentclass[twocolumn,notitlepage,prl,superscriptaddress,longbibliography]{revtex4-2}
\usepackage{amsfonts}
\usepackage{textcomp}
\usepackage{times}
\usepackage{graphicx}
\usepackage{float}
\usepackage{latexsym,amsmath,amssymb,bm,euscript}
\usepackage{color}
\usepackage{subfigure}
\usepackage{epstopdf}
\usepackage[colorlinks=true,linkcolor=blue,citecolor=blue]{hyperref}
\usepackage{soul}
\usepackage[normalem]{ulem}
\usepackage{mathrsfs}
\usepackage{amsmath}
\usepackage{xspace}
\usepackage{natbib}
\usepackage{ulem}
\usepackage{etoolbox}
\usepackage{tikz}
\usepackage{physics}
\usepackage{chemformula}
\usepackage{natbib}
\usepackage{tikz}
\usepackage{makecell} 

\def\nbwo{Nd$_3$BWO$_9$\xspace}

\newcommand{\nbcp}{\ch{Na_2BaCo(PO_4)_2}}
\newcommand{\cmn}{Ce$_2$Mg$_3$(NO$_3$)$_{12}\cdot$24H$_2$O}

\begin{document}
\title{Ising Supercriticality and Universal Magnetocalorics in Spiral Antiferromagnet \nbwo}

\author{Xinyang Liu}
\thanks{These authors contributed equally to this work.}
\affiliation{School of Physics, Beihang University, Beijing 100191, China}
\affiliation{Beijing National Laboratory for Condensed Matter Physics, Institute of Physics, Chinese Academy of Sciences, Beijing 100190, China}
\affiliation{Peng Huanwu Collaborative Center for Research and Education, Beihang University, Beijing 100191, China}

\author{Enze Lv}
\thanks{These authors contributed equally to this work.}
\affiliation{Institute of Theoretical Physics, Chinese Academy of Sciences, Beijing 100190, China}
\affiliation{School of Physical Sciences, University of Chinese Academy of Sciences, Beijing 100049, China}

\author{Xueling Cui}
\thanks{These authors contributed equally to this work.}
\affiliation{School of Physics, Beihang University, Beijing 100191, China}

\author{Han Ge}
\thanks{These authors contributed equally to this work.}
\affiliation{Department of Physics, Southern University of Science and Technology, Shenzhen, China}

\author{Fangyuan Song}
\affiliation{School of Physics and Wuhan National High Magnetic Field Center, Huazhong University of Science and Technology, Wuhan 430074, China}

\author{Zhaoming Tian}
\affiliation{School of Physics and Wuhan National High Magnetic Field Center, Huazhong University of Science and Technology, Wuhan 430074, China}

\author{Gang Su}
\affiliation{Institute of Theoretical Physics, Chinese Academy of Sciences, Beijing 100190, China}

\author{Kan Zhao}
\email{kan\_zhao@buaa.edu.cn}
\affiliation{School of Physics, Beihang University, Beijing 100191, China}

\author{Junsen Xiang}
\email{xiangjs@iphy.ac.cn}
\affiliation{Beijing National Laboratory for Condensed Matter Physics, Institute of Physics, Chinese Academy of Sciences, Beijing 100190, China}

\author{Peijie Sun}
\email{pjsun@iphy.ac.cn}
\affiliation{Beijing National Laboratory for Condensed Matter Physics, Institute of Physics, Chinese Academy of Sciences, Beijing 100190, China}

\author{Wei Li}
\email{w.li@itp.ac.cn}
\affiliation{Institute of Theoretical Physics, Chinese Academy of Sciences, Beijing 100190, China}
\affiliation{Peng Huanwu Collaborative Center for Research and Education, Beihang University, Beijing 100191, China}
\date{\today}

% =============================================================
%                          ABSTRACT
% =============================================================
\begin{abstract}
The celebrated analogy between the pressure-temperature phase diagram of a liquid-gas system and the field-temperature phase diagram of a ferromagnet has long been a cornerstone for understanding universality of phase transitions and critical phenomena. Here we extend this analogy to a highly frustrated antiferromagnet, the spiral Ising compound \nbwo~with kagome layers. In its phase diagram, we identify a metamagnetic transition line with a critical endpoint (CEP) located at $\mu_0H_{\mathrm{c}} \simeq 1.04$ T and $T_{\mathrm{c}} \simeq 0.3$ K. Above the CEP, an Ising supercritical regime emerges with crossover lines that follow a universal scaling law, as evidenced by the specific heat, magnetic susceptibility, and magnetocaloric measurements. Remarkably, we observe highly sensitive magnetic cooling near the emergent CEP, characterized by a divergent magnetic Gr\"uneisen ratio $\Gamma_H \propto 1/t^{\beta+\gamma-1}$, with $\beta + \gamma \simeq 1.563$ the sum of critical exponents of the 3D Ising universality class and $t \equiv (T-T_{\rm c})/T_{\rm c}$ the reduced temperature. Adiabatic demagnetization from 2~K and 4~T reaches a minimum temperature of 195~mK, via a self-cascading process that combines supercritical and topological cooling. Our findings open a new avenue for studying supercritical phenomena and magnetic refrigeration with the frustrated rare-earth compounds RE$_3$BWO$_9$ and, more broadly, in Ising‑anisotropic antiferromagnets such as spin ices.
\end{abstract}
\maketitle

% ==================================================================
%                             INTRODUCTION
% ==================================================================
\textit{Introduction.---}
The critical endpoint (CEP), beyond which the distinction between liquid and gas vanishes, has long served as a landmark concept in phase transitions and critical phenomena~\cite{Cagniard1822}. At the CEP, the first-order transition line terminates, and the system exhibits scale invariance and an emergent global $\mathbb{Z}_2$ symmetry~\cite{li2024supfluid}. The resulting scaling laws place the liquid-gas transition in the three-dimensional (3D) Ising universality class, establishing a profound correspondence with ferromagnetic (FM) phase transitions~\cite{Lee1952}. Above the CEP, the system enters a strongly fluctuating supercritical regime that has found broad applications ranging from refrigeration and engines to power generation~\cite{Ackerman1970JHT, Clifford2000, Edet2023, KIM2004SCR}. The emergent CEP and supercritical phenomena have recently attracted renewed interest in the context of strongly correlated quantum materials~\cite{Mila2021, Wang2023, Tang2022NP, wang2024qsc, Lv2025qsr, Wu2025, Sylvia2025}.

Frustrated spin systems offer an ideal platform for uncovering exotic magnetic states and transitions~\cite{Sachdev2008QMag, Sachdev2000, Balents2010, Sachdev2015, Zhou2017RMP, Broholm2020}. Among them, rare-earth magnets are particularly compelling, in which the interplay of geometric frustration and quantum fluctuations can give rise to a rich landscape of exotic spin states~\cite{Wu2019YAO, Kish2025YAO, Shen2016, Han2020, Hu2020, Zhou2024RESSL, Liu2024YBGO}. Recently, a new family of rare-earth compounds with stacked kagome layers, RE$_3$BWO$_9$ (RE = Pr-Sm, Gd-Ho), has been discovered~\cite{Krutko2006lz, Tian2020IC}. In particular, the compound \nbwo\ hosts \ch{Nd^{3+}} ions that form spin-orbit Kramers doublets, constituting an effective $S = 1/2$ frustrated antiferromagnet~\cite{AZ2023PRB, Tian2023PRB}. Initially, its kagome-layered structure was thought to potentially host a quantum spin liquid phase, sparking significant research interest~\cite{Tian2020IC, AZ2023PRB, Tian2023PRB, AZ2025PRBL, Khuntia2025PRB}.

% =====================================
%                             FIG 1
% =====================================
\begin{figure}[t!]
\centering
\includegraphics[width=0.88\linewidth]{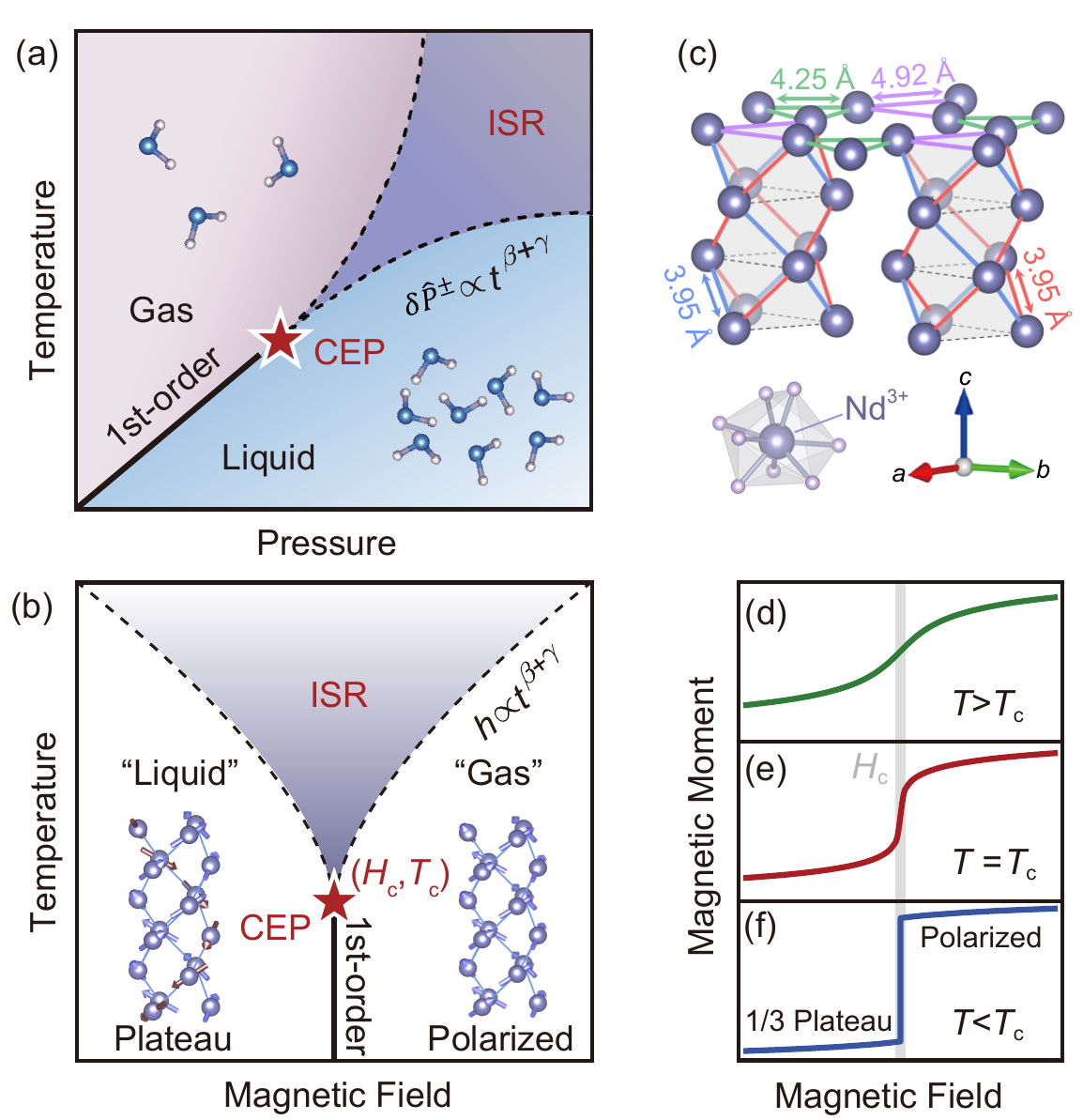}
\caption{
Schematic phase diagrams of (a) water and (b) \nbwo, where the solid line represents a first-order transition and the red star denotes the critical endpoint (CEP). The dashed lines enclose the Ising supercritical regime (ISR). The universal supercritical crossover scaling is $\delta \hat{P}^{\pm}\propto t^{\beta+\gamma}$ for water~\cite{li2024supfluid}, and $h\propto t^{\beta+\gamma}$ for \nbwo\ ($\delta\hat{P}^{\pm}$, $t$ and $h$ being reduced parameters). $\beta$ and $\gamma$ are the critical exponents of the 3D Ising universality class. (c) Coupled Ising tubes formed by Nd$^{3+}$ ions and distorted NdO$_8$ octahedra. (d-f) Schematic magnetization curves above, at, and below $T_{\rm c}$. 
}
\label{fig1}
\end{figure}

% ===========================
%                         FIG 2
% ===========================
\begin{figure*}[t!]
\centering
\includegraphics[width=1\linewidth]{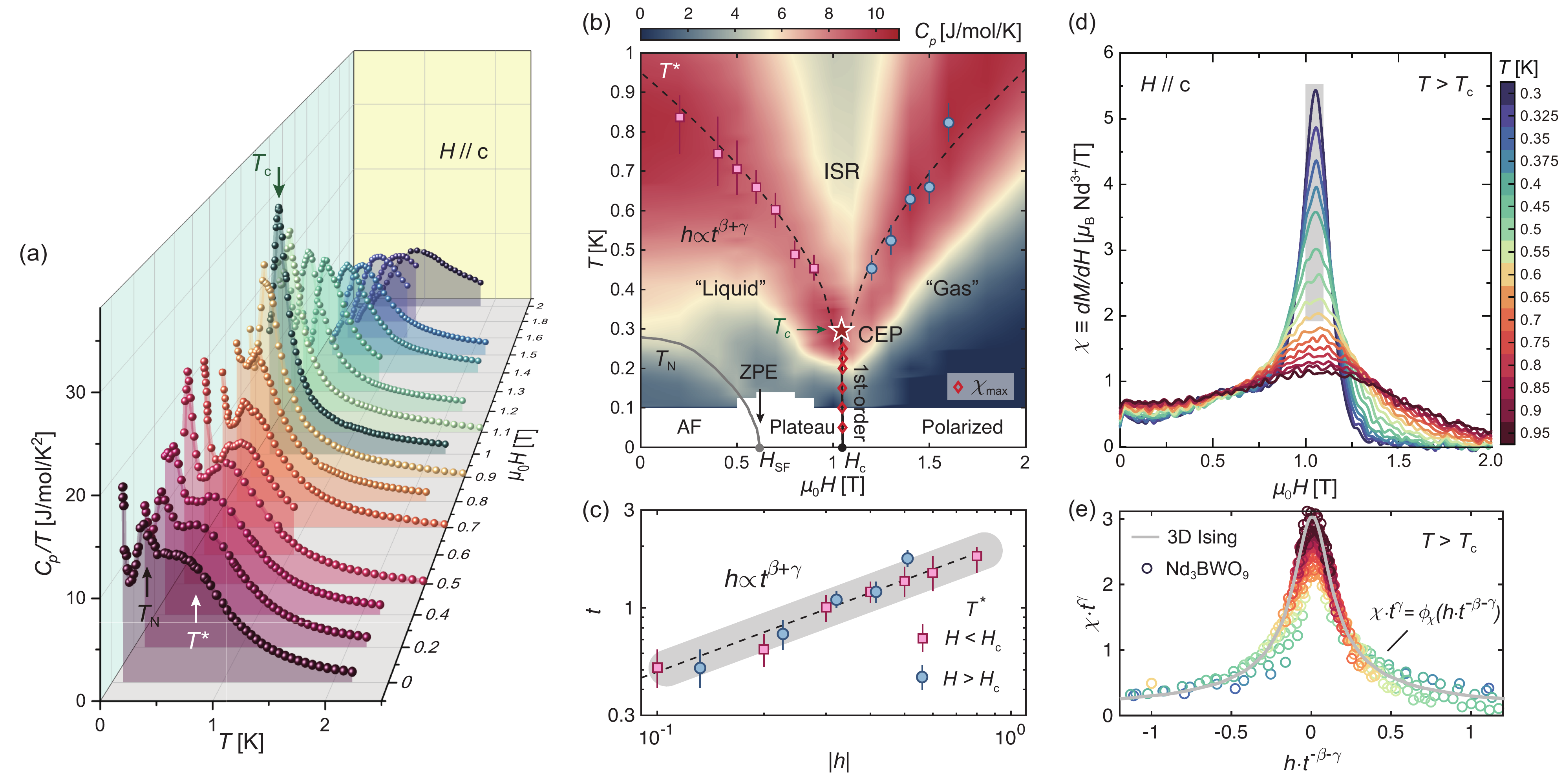}
\caption{
(a) Low-temperature specific heat ($C_p/T$) of \nbwo\ under $c$-axis magnetic fields from 0 T to 2 T.
(b) Color map of $C_p$, where the maxima for $H < H_{\rm c}$ (red squares) and $H > H_{\rm c}$ (blue circles) denote the supercritical crossovers (dashed black lines). The uncertainty (error bar) is estimated by the temperature step size in the specific heat measurements. The star marks the CEP at $\mu_0H_{\rm c} \simeq 1.04(4)$~T and $T_{\rm c} \simeq 0.30(2)$ K. The uncertainties reflect the variation across different measurements, including the specific heat, magnetization, and magnetocaloric measurements. The solid black line below the CEP represents the first-order metamagnetic transition line. The red diamonds mark the peak positions of $\chi \equiv dM/dH$ below $T_{\rm c}$, which are used to identify the first-order line~\cite{SM}. Above the CEP, there exists an ISR which separates the ``liquid'' and ``gas'' states. The gray dot indicates the spin-flip field $\mu_0H_{\rm SF} \simeq 0.65$~T, and the solid gray line is the AF phase boundary. For $H_{\rm SF}<H<H_{\rm c}$, we find no finite-$T$ phase transitions but crossovers.
(c) Supercritical crossover scaling law $h\propto t^{\beta+\gamma}$, with $\beta$ and $\gamma$ the critical exponents of the 3D Ising universality class. The reduced field is $h \equiv (H-H_{\rm c})/H_{\rm c}$, and the reduced temperature is $t \equiv (T^*-T_{\rm c})/T_{\rm c}$, where $T^{*}$ denotes the peak positions of $C_p$ in panel (b). 
(d) Magnetic susceptibility $\chi$ as a function of magnetic field at various temperatures above $T_{\rm c}$.
(e) Data collapse of the measured $\chi$ results in the supercritical regime, whose profile is in excellent agreement with the universal scaling function $\phi_{\chi}(x)$ obtained theoretically,  from the 3D Ising model calculation (see End Matter). Data in the gray region are omitted from the data collapse due to the limited experimental resolution. In this highly field-sensitive regime, the diverging $\chi$ is cut off by the finite field step of 0.02~T. 
}
\label{fig2}
\end{figure*}

% ====================================================================================
%                                      FIG 3
% ====================================================================================
\begin{figure*}[t!]
\centering
\includegraphics[width=1\linewidth]{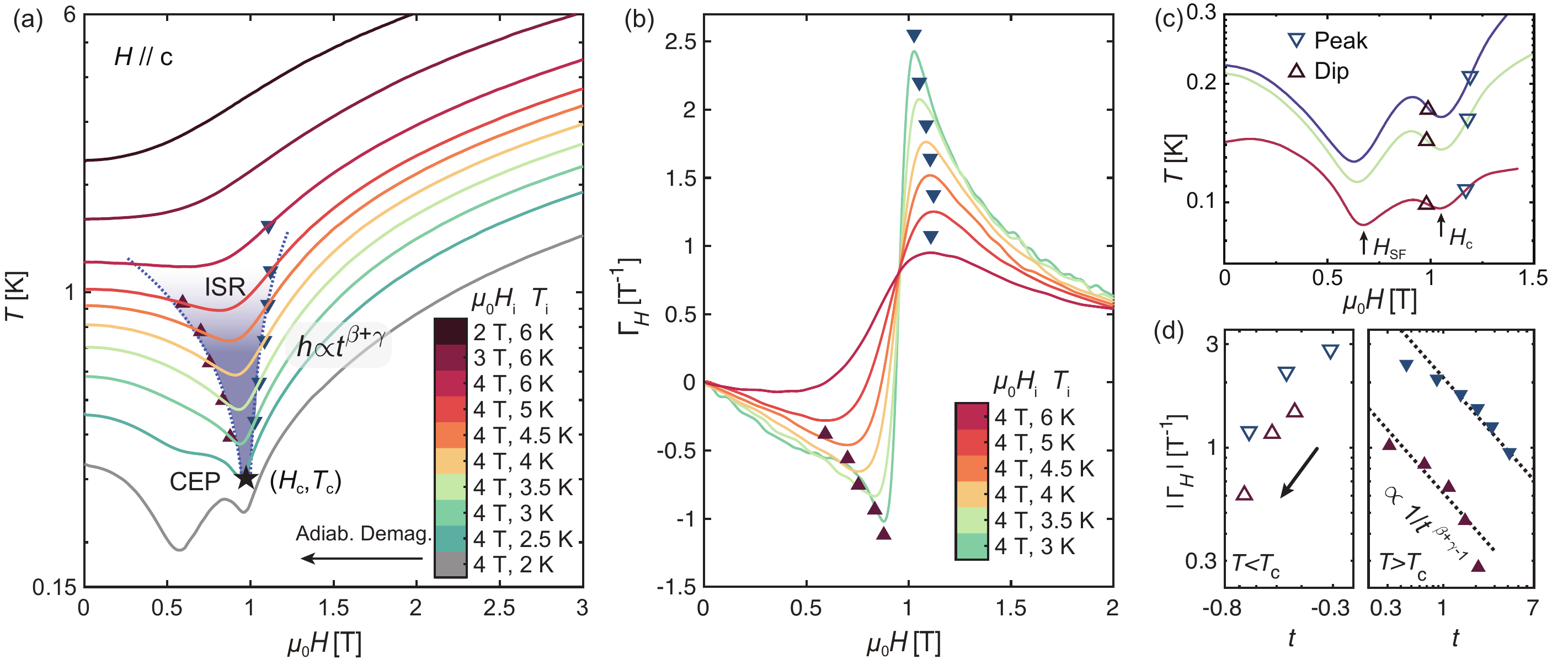}
\caption{
(a) The isentropes obtained via adiabatic demagnetization measurements, with legends specifying the initial conditions ($H_{\rm i},T_{\rm i}$). The supercritical crossover line (blue dotted line) $h \propto t^{\beta+\gamma}$ is determined from the peak/dip positions of the magnetic Gr\"uneisen ratio $\Gamma_H$ [panel (b)], derived from the isentropic lines.
(b) Magnetic Gr\"uneisen ratio $\Gamma_H\equiv \frac{1}{T}\left(\frac{\partial T}{\partial H}\right)_S$, where blue and red triangles mark the peaks and dips, respectively.
(c) Isentropic lines below $T_{\rm c} \simeq 0.3$ K. Hollow triangles mark the peak/dip positions of the corresponding $\Gamma_H$.
(d) Temperature dependence of the peak/dip values of $\Gamma_H$. The black dashed line indicates the supercritical magnetocaloric scaling $\Gamma_H \propto 1/t^{\beta+\gamma-1}$ for $T > T_{\rm c}$. Very close to $T_{\rm c}$, deviations from this scaling arise from limited measurement resolution. On the subcritical side ($T < T_{\rm c}$), the extrema decrease rapidly. 
}
\label{fig3}
\end{figure*}

Recently, neutron scattering studies have established an antiferromagnetic (AF) order in \nbwo\ below 0.3~K~\cite{AZ2023PRB}, ruling out a spin-liquid ground state. Subsequent measurements revealed a complex spin order due to the triple braids of spiral Ising axes~\cite{AZ2025PRBL}. The AF, 1/3-plateau, and partially polarized states are separated by two first-order spin state transitions due to the strong magnetic anisotropy. A magnetization jump occurs at the metamagnetic transition from the 1/3-plateau to the partially polarized phase [see Fig.~\ref{fig1}(b)]~\cite{AZ2025PRBL, Tian2023PRB}. This jump vanishes above a critical temperature, as shown schematically in Figs.~\ref{fig1}(d-f); however, anomalies persist in the specific heat at higher temperatures~\cite{AZ2023PRB, Tian2023PRB}. A rich field-temperature phase diagram has been mapped out by magnetic and thermodynamic measurements, complemented by low-energy dynamical probes of spin fluctuations via $\mu$SR and NMR~\cite{Khuntia2025PRB}. These measurements collectively establish \nbwo\ as a compelling platform for investigating field-induced magnetic transitions and critical phenomena.

Here we report the discovery of a field-driven supercritical regime in the spiral Ising antiferromagnet \nbwo, emerging above a CEP on the metamagnetic first-order transition line. In the phase diagram of \nbwo, there exists a field-induced first-order metamagnetic line that terminates at the CEP with $\mu_0H_{\rm c} \simeq 1.04(4)$ T and $T_{\rm c} \simeq 0.30(2)$~K. Above the CEP, there exists a field-induced Ising supercritical regime (ISR), which hosts strongly fluctuating spin states highly sensitive to external fields. As illustrated in Fig.~\ref{fig1}(b), we find the crossover lines adhere to the supercritical scaling $h \propto t^{\beta + \gamma}$ with $\beta+\gamma \simeq 1.563$ of the 3D Ising universality class~\cite{li2024supfluid, 3DIsing2025}. This scaling manifests as two ``ridges'' in the specific heat emanating from the CEP [Figs.~\ref{fig2}(a,b)] and is further corroborated by a collapse of the magnetic susceptibility onto the universal function $\phi_{\chi}(x)$. Our findings thus establish an antiferromagnetic analog of liquid-gas Ising supercriticality.
% { the liquid and gas states become indistinguishable. As shown in Fig.~\ref{fig1}{(a)}, the system enter a strongly fluctuating Ising supercritical regime (ISR) and the crossover lines that enclose the supercritical fluid follow a supercritical scaling~\cite{li2024supfluid}. In \nbwo, we also identify a supercritical state [see Fig.~\ref{fig1}{(b)}], which hosts strong thermal fluctuations and is sensitive to external magnetic field.}

Remarkably, the high field-sensitivity near the CEP gives rise to a supercritical magnetocaloric effect (MCE), characterized by the universally diverging Gr\"uneisen ratio, $\Gamma_H \equiv 1/T (\partial T/\partial H)_S \propto 1/t^{\beta + \gamma - 1}$. Through adiabatic demagnetization measurements on \nbwo, we cool the sample to 195 mK near the lower spin-flip field ($\mu_0 H_{\rm SF} \simeq 0.65$ T), enabled by both supercritical and topological cooling. The latter can be ascribed to a proximate zero-point entropy of domain-wall defects. Our results establish the boratotungstate \nbwo\ as an efficient sub-Kelvin coolant with exceptionally high spin density. These findings may find broad applicability in other magnetic systems with strong Ising anisotropy and local constraints like spin-ice compounds~\cite{Ramirez1999Nature, Aoki2004JPSJ, Morris2009Science, Pomaranski2013NP, Steven2001Science, Castel2008Nature, Tang2022NP, Wen2017PRL, Thompson2017PRL, Gaudet2019PRL}.

% =======================================
%                               Results
% =======================================
\textit{Universal thermal data with supercritical scaling.---}
We synthesize high-quality \nbwo single crystals (see End Matter) and measure the specific heat $C_p/T$ under $c$-axis magnetic fields down to 100 mK [see Fig.~\ref{fig2}(a)]. Under zero field, the specific heat exhibits a sharp peak at about $T_N \simeq 0.29$ K, signaling the onset of AF order. With increasing field, this peak shifts to lower temperatures and vanishes at $\mu_0 H_{\rm SF} \simeq 0.65$~T. Notably, a broad hump observed at $T^* \approx 1$~K, which marks the buildup of short-range spin correlations, shifts downward with increasing field. The hump position reaches a minimum temperature of about 0.3~K near $H_{\rm c}$ before turning upward at higher fields. 

Near the CEP, locations of the humps trace the boundaries of the finite-temperature ISR, and obey the universal scaling law $h \propto t^{\beta+\gamma}$ of the 3D Ising universality class [see Fig.~\ref{fig2}(c)]. Such scaling behavior originates from the universal form of specific heat, i.e., $C(H,T) = t^{-\alpha} \phi_C(x)$, where $\phi_C(x)$ with $x\equiv h \cdot t^{-\beta-\gamma}$ is the scaling function derived from the singular part of the free energy (see End Matter). The order parameter for the first-order transition between the 1/3-plateau and partially polarized phases is the uniform magnetization. This is analogous to the density difference across the first-order liquid-gas phase boundary. As in the liquid-gas system, the CEP here emerges without spontaneous symmetry breaking yet exhibits an emergent $\mathbb{Z}_2$ symmetry, owing to the low symmetry of the spin structures under magnetic fields~\cite{Tian2023PRB}. Moreover, while the emergent CEP is confined to a single point, the ISR, where ``liquid'' and ``gas'' are indistinguishable, extends over a broad region, making it far more accessible in experiments.

In addition, we conduct ultralow-temperature magnetization measurements to determine the metamagnetic transition at $\mu_0H_{\rm c} \simeq 1.04$~T from the magnetization jump~\cite{SM}. The magnetic susceptibility can be obtained from the magnetization measurements, by differentiation $\chi \equiv dM/dH$. We show the supercritical data in Fig.~\ref{fig2}(d), where the series of sharp peaks near $H_{\rm c}$ indicates the high field-sensitivity; below $T_{\rm c}$, the divergence of $\chi$ is cut off by a finite step length in the $M$-$H$ measurements. According to the scaling analysis, magnetic susceptibility follows a universal scaling form $\chi(H,T) = t^{-\gamma} \phi_{\chi}(h \cdot t^{-\beta-\gamma})$ above the CEP. By rescaling $x \equiv  h \cdot t^{-\beta -\gamma}$ and $y \equiv \chi \cdot t^{\gamma}$ with the critical exponents $\beta \simeq 0.326$ and $\gamma \simeq 1.237$, we collapse the magnetic susceptibility data onto a single curve [see Fig.~\ref{fig2}(e)]. Notably, it is in excellent agreement with the calculated scaling function $\phi_{\chi}(x)$ for the 3D Ising universality class (see End Matter). %Such a behavior is consistent with a CEP at finite $T_{\rm c}$ rather than a quantum critical scenario.

% ====== Supercritical MCE ====== %
\textit{Supercritical magnetocalorics {and ultralow-$T$ cooling}.---} 
The adiabatic demagnetization measurements of \nbwo starting from different initial conditions yield its isentropic lines. As shown in Fig.~\ref{fig3}(a), isentrope dips appear within the ISR, due to the strong fluctuations of supercritical spin states. The corresponding magnetic Gr\"uneisen ratio $\Gamma_{H}$, derived from the isentropic lines, displays a peak-dip structure [see Fig.~\ref{fig3}(b)]. As the temperature approaches $T_{\rm c}$, the peak and dip positions asymptotically approach $H_{\rm c}$. Moreover, in Fig.~\ref{fig3}(d), we find the peak/dip values obey a supercritical magnetocaloric scaling law $\Gamma_H \propto 1/t^{\beta+\gamma-1}$, which can also be derived from the universal form of free energy (see End Matter). Below $T_{\rm c}$, the isentrope dips near $H_{\rm c}$ grow progressively shallower at lower temperatures, as evidenced in Fig.~\ref{fig3}(c). This reflects the weakened cooling effect due to the minimum entropy change of the metamagnetic transitions.

\begin{figure}[t!]
\centering
\includegraphics[width=0.95\linewidth]{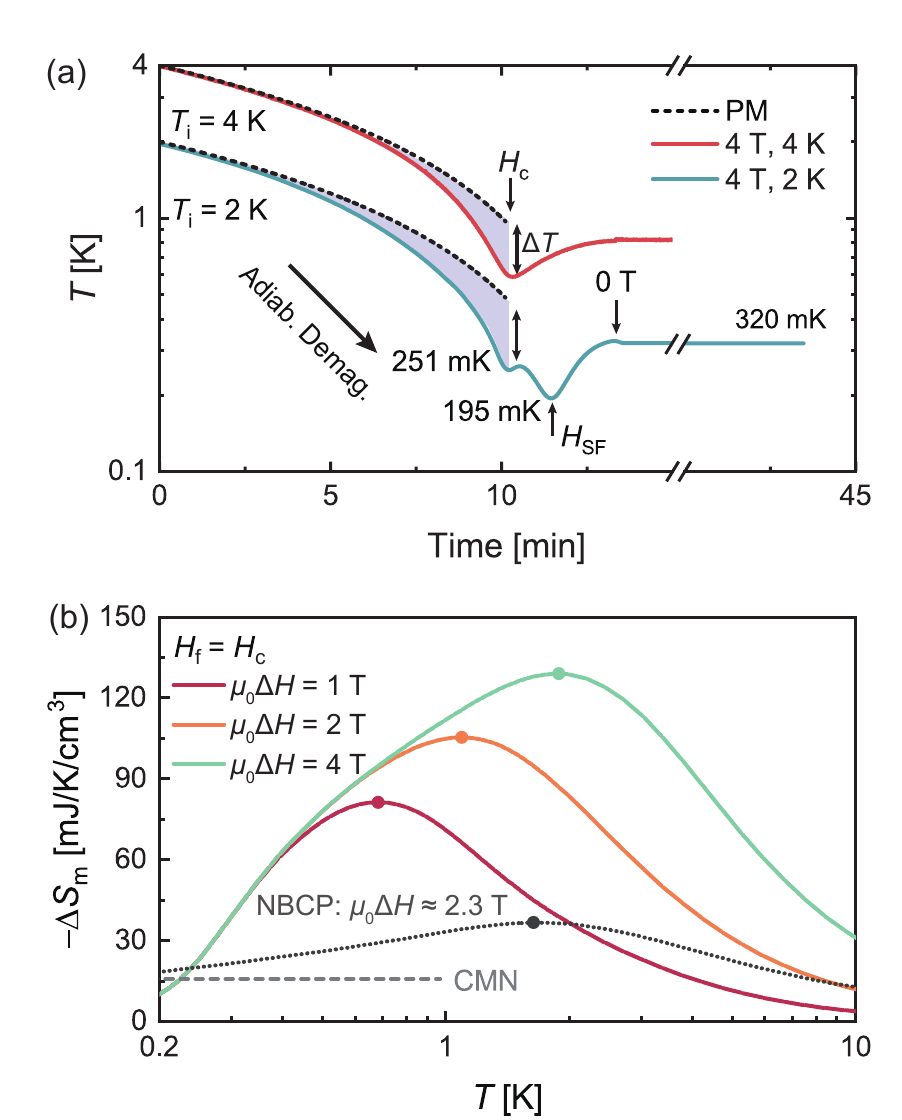}
\caption{
(a) Cooling curves during adiabatic demagnetization under fields along the $c$ axis. The green and red solid lines trace the temperature evolution from the same initial field $\mu_0H_{\rm i} = 4$ T and different initial temperatures $T_{\rm i} = 4$ K and 2 K, respectively. The black dashed lines represent the cooling curves of an ideal paramagnetic (PM) salt. The purple shaded areas highlight the enhanced cooling effect in \nbwo\ compared to the PM salt, characterized by $\Delta T \simeq 414$~mK and 249~mK for $T_{\rm i} \simeq 4$ K and 2 K, respectively.
(b) Volumetric magnetic entropy change ($-\Delta S_{\rm m}$) for \nbwo\ under different field changes (final field at $\mu_0H_{\rm f} = \mu_0H_{\rm c}\simeq 1$~T). The dots mark the maximum of each curve. The black dotted line shows the spin supersolid compound \nbcp~(NBCP) under a field change from 4 T to its critical field 1.66~T~\cite{Xiang2024Nature}; and the gray dashed line shows the {zero-field entropy} of the prototypical PM coolant \cmn~(CMN). 
}
\label{fig4}
\end{figure}

Figure~\ref{fig4}(a) illustrates two representative adiabatic demagnetization refrigeration measurements starting from $(\mu_0H_{\rm i},\,T_{\rm i}) = (4\,{\rm T},\, 4\,{\rm K})$ and $(4\,{\rm T},\, 2\,{\rm K})$, respectively. For the process starting from $(4\,{\rm T},\, 4\,{\rm K})$, \nbwo cools to 586~mK near $H_{\rm c}$, surpassing the ideal paramagnetic (PM) salt (dashed lines); starting from $(4\,{\rm T},\, 2\,{\rm K})$, it reaches the lowest temperature of 195~mK at the spin-flip field $\mu_0 H_{\rm SF} \simeq 0.65$~T [see also Fig.~\ref{fig3}(a,c)]. The lower cooling temperature reflects enhanced low-$T$ entropy near $H_{\rm SF}$, which shifts the temperature minimum to the spin-flip field.

A key figure of merit for a magnetic refrigerant is its volumetric entropy change $-\Delta S_{\rm m}\equiv S_{\rm m}(H_{\rm f},T) - S_{\rm m}(H_{\rm i},T)$, which largely determines practical cooling capacity. For \nbwo, we calculated $-\Delta S_{\rm m}$ for each field change from the entropy data (see End Matter). For a modest field change of 1~T (and stopping at $H_{\rm f}$ = $H_{\rm c}$), we find a large entropy change of $-\Delta S_{\rm m} \simeq 83$ mJ$\cdot$K$^{-1}\cdot$cm$^{-3}$ in the sub-Kelvin regime [see Fig.~\ref{fig4}(b)]. This is substantially higher than that of NBCP, even for smaller field windows, reflecting the high field-sensitivity of supercritical MCE. The entropy change value of \nbwo also far exceeds that of the spin-1/2 hydrated PM salts, attributed partly to its exceptionally high spin density ($N \simeq 16.9$ nm$^{-3}$), an order of magnitude greater than that of the hydrate CMN~($N \simeq 1.65$ nm$^{-3}$). 

% =====================================
%                                      FIG 4
% =====================================

\textit{Proximate zero-point entropy in spiral Ising tube.---}
To interpret these features, we perform model calculations involving a frustrated spiral Ising tube, as spin couplings in the kagome layer are negligibly weak~\cite{AZ2025PRBL} (see End Matter). We uncover a first-order spin-flip transition at $H_{\rm SF}$ carrying a macroscopic ground-state degeneracy. This leads to a proximate zero-point entropy (ZPE), calculated as $S_0 \simeq 0.481 R$ per unit cell and evidenced by the shoulder in the zero-field entropy curves (Fig.~\ref{figE3}). This extensive degeneracy can be ascribed to the presence of magnetic domain walls --- topological defects that correspond to global spin flips throughout the spiral Ising tube. Therefore, starting from $(4\,{\rm T},\, 2\,{\rm K})$, a two-step cooling process occurs: one driven by supercritical fluctuations near $H_{\rm c}$ and the other by proximate ZPE near $H_{\rm SF}$. It is such a self-cascading mechanism, i.e., two successive cooling dips in a single demagnetization sweep, in \nbwo that achieves an ultralow temperature of 195~mK.

Furthermore, a comparison with FM coolants of comparable spin density, such as \ch{LiHoF_4} ($N \simeq 13.7$ nm$^{-3}$), reveals that while the latter also exhibits high field-sensitivity, its ordering temperature is significantly higher ($T_{\rm c} \simeq 1.53$ K)~\cite{Xie2021Giant, Wendl2022Mesoscale, Liu2023NPG}. This highlights a key distinction: the CEP in \nbwo is an emergent phenomenon whose $T_{\rm c} \simeq 0.3$~K  is greatly suppressed by spin frustration --- despite the very high ion density and a significant spin coupling of about 3 K. This underscores the important role of magnetic frustration in this spiral Ising antiferromagnet. Although kagome-plane couplings are negligible due to nearly orthogonal Ising axes~\cite{AZ2025PRBL}, frustration in \nbwo stems from competing FM-AF interactions along the three braided chains. 

% ===============================================
%                           CONCLUSION AND OUTLOOK
% ===============================================
\textit{Discussion.---} 
In this work, we report comprehensive thermodynamic and magnetocaloric measurements, with scaling analysis of the supercritical behaviors, in \nbwo. We achieve the lowest temperature of 195~mK through self-cascading cooling between two field-induced metamagnetic transitions: one at $\mu_0 H_{\rm c} \simeq 1.04$~T associated with a low-$T_{\rm c}$ CEP and the other at $\mu_0 H_{\rm SF} \simeq 0.65$~T  arising from proximate ZPE. Whereas the MCE has traditionally been explored in FM metals~\cite{Weiss1917} and, more recently, in FM insulators with low Curie temperatures~\cite{Vilches1966FAA, DAUDIN1982GGG, Liang2015NC, Xie2021Giant, Liu2023NPG, Mo2024Jacs, Mo2025JACS}, our work exploits highly frustrated, Ising-anisotropic antiferromagnets harboring a low-$T_{\rm c}$ CEP and a highly field-sensitive Ising supercritical regime. The universal scaling behaviors unveiled near the Ising CEP suppressed by spin frustration, together with the associated supercritical MCE, offer valuable insights into other members of the kagome-layered RE$_3$BWO$_9$ family~\cite{Pi2021PRB, Zeng2022CPL, AZ2024PRR, Song2025CPL, Wang2025PRM}, such as Sm$_3$BWO$_9$ with its Kramers doublet and an incommensurate spin order along the $c$ axis~\cite{Zeng2022CPL}.

Furthermore, we emphasize a profound similarity between the phase diagram of \nbwo\ and those of spin-ice systems, which share strong Ising anisotropy and proximate zero-point entropy --- a hallmark of spin-ice physics. Both classical (e.g., Dy$_2$Ti$_2$O$_7$~\cite{Ramirez1999Nature, Aoki2004JPSJ, Pomaranski2013NP, Morris2009Science}) and quantum (e.g., Pr$_2$Zr$_2$O$_7$~\cite{Wen2017PRL, Tang2022NP}) spin ices host first-order metamagnetic transition lines terminating at a CEP, above which a supercritical regime emerges. The spin-ice compound \ch{Dy_2Ti_2O_7}, for example, possesses a comparable spin density ($N \simeq 15.44$ nm$^{-3}$) and a low-$T_{\rm c}$ CEP at 0.36~K~\cite{Aoki2004JPSJ}, whereas Pr$_2$Zr$_2$O$_7$ exhibits an even lower $T_{\rm c} \simeq 0.06$~K~\cite{Tang2022NP, SM}. These compounds may likewise prove to be excellent sub-Kelvin supercritical coolants.  Given the global shortage of helium-3~\cite{Kramer2019Helium, Cho2009Science}, the search for advanced magnetic refrigerants has intensified~\cite{Xiang2024Nature, Shu2026Nature}. Our findings establish the Ising supercriticality in frustrated antiferromagnets as a novel mechanism for ultralow-temperature magnetic refrigeration.

% ================================================================
%                       ACKNOWLEDGMENTS
% ================================================================

\begin{acknowledgments}
\textit{Acknowledgments.---}
E.L. and W.L. are grateful to Yuan Gao, Junsen Wang, Yang Qi, and Yuliang Jin for insightful discussions. K.Z. extends thanks to Xingye Lu for valuable experimental assistance. This work was supported by the National Key Projects for Research and Development of China (Grant Nos.~2024YFA1409200, 2023YFA1406003, and 2022YFA1402200), the National Natural Science Foundation of China (Grant Nos.~12534009, 12447101, 12404180, 12274015, and 12474147), the Strategic Priority Research Program of Chinese Academy of Sciences (Grant No.~XDB1270100), the Beijing Natural Science Foundation (Grant No.~JQ24012), and the Fundamental Research Funds for the Central Universities. 
We thank HPC-ITP for the technical support and generous allocation of CPU time. We acknowledge the support from the Synergetic Extreme Condition User Facility (SECUF, https://cstr.cn/31123.02.SECUF), and the facilities and technical support of the Extreme Condition Characterization Platform at Analysis \& Testing Center of Beihang University.
\end{acknowledgments}

\textit{Data availability.---}
The data that support the findings of this work are openly available \cite{Database}.

\bibliography{nbwo.bib}

% ===================================
%                                End Matter
% ===================================

\clearpage
\newpage
\onecolumngrid
\begin{center}
\textbf{\large{End Matter}}
\end{center}
\twocolumngrid

\begin{figure}[h!]
\centering
\includegraphics[width=0.75\linewidth]{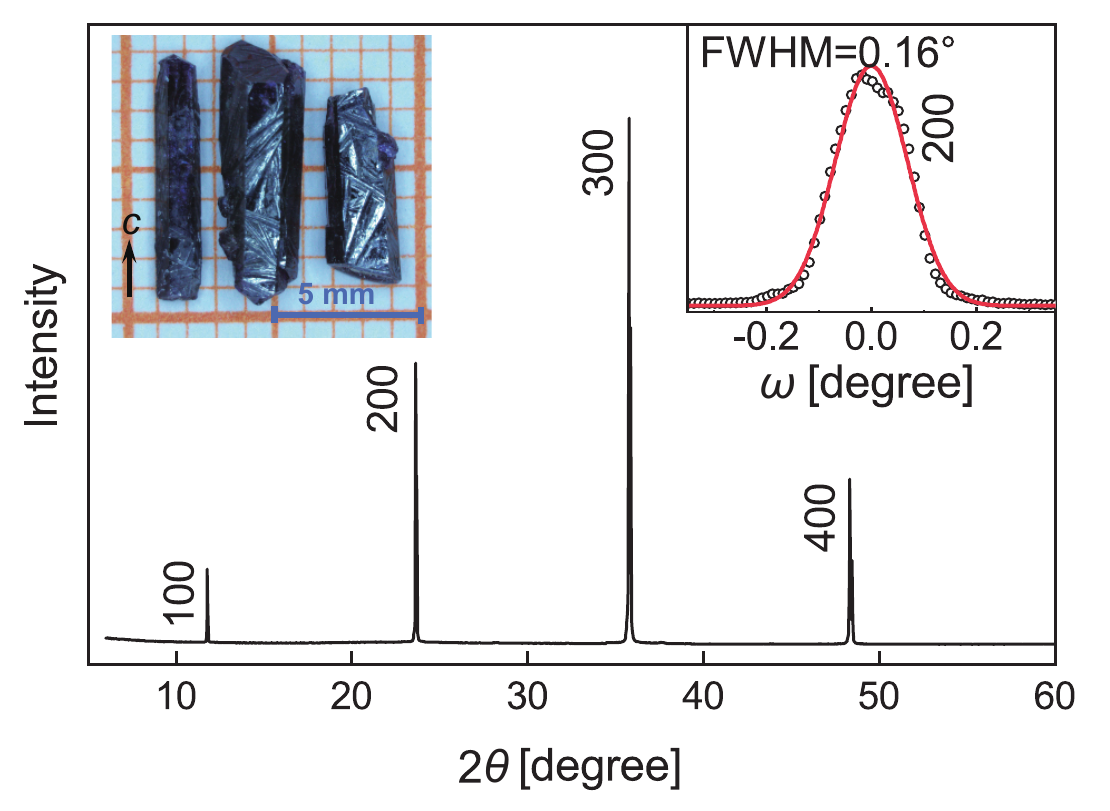}
\caption{X-ray diffraction pattern of \nbwo single crystal. Left inset: an optical image of typical single crystals. Right inset: the rocking curve (black line) of the (200) Bragg peak, together with the Gaussian fitting (red line). 
}
\label{figE1}
\end{figure}

\textit{Sample synthesis and characterization.---} 
Single crystals of \nbwo were synthesized using the PbO flux method. A mixture of \nbwo powder and PbO in a 1:14 molar ratio was thoroughly ground and placed in a platinum crucible. The crucible was heated to 1150 $^\circ$C, held for 25 h, cooled to 900 $^\circ$C at a rate of 2.5 $^\circ$C/h, and finally cooled to room temperature at 100 $^\circ$C/h. Crystals were separated by etching in hot acid, as shown in the left inset of Fig.~\ref{figE1}. The X-ray diffraction pattern of the $bc$ plane displays dominant (H00) reflections at room temperature, using a Bruker D8 ADVANCE diffractometer with Cu-K$\alpha$ radiation ($\lambda = 1.5406$ Å). Rocking curve analysis of the Bragg peak (200), right inset of Fig.~\ref{figE1}, demonstrates a narrow full width at half maximum (FWHM) of 0.16°, indicating high crystal quality. 

% ====================================================================================
%                                      FIG E1
% ====================================================================================

\textit{Thermodynamic and magnetocaloric measurements.---}
Specific heat measurements above 2 K are performed using the standard heat capacity option of the Physical Property Measurement System (PPMS). Ultralow-temperature specific heat measurements are carried out using the quasi-adiabatic heat-pulse method, and magnetization measurements down to 50 mK are performed using a Hall-sensor magnetometer, both implemented in a PPMS equipped with a $^3$He-$^4$He dilution refrigerator.

Cooling curves via adiabatic demagnetization are measured using a custom-designed setup based on the PPMS. Thermal insulation is achieved by employing low thermal-conductivity support structures and a high-vacuum environment. Multiple single crystals of \nbwo (total mass 1.5~g) are aligned in the same direction and used as samples, and their temperature is monitored using a calibrated RuO$_2$ thermometer. The magnetic field is applied along the $c$ axis, and demagnetization is carried out at a constant sweep rate of 50~Oe/s.

\textit{Scaling analysis of universal thermodynamics.---}
Above the critical endpoint (CEP), the singular part of the free energy possesses a universal form as 
\begin{equation}
    F=t^{2-\alpha} \xi_1 \phi_{F}(\xi_2 ht^{-\beta-\gamma}),\,{\rm for}\, t>0,
    \label{Equ:FreeEnergy}
\end{equation} 
where $t\equiv (T-T_{\rm c})/T_{\rm c}$ and $h\equiv (H-H_{\rm c})/H_{\rm c}$ are reduced parameters measuring the distance to the CEP at $(H_{\rm c}, T_{\rm c})$. $\beta$ and $\gamma$ are critical exponents. $\xi_1$ and $\xi_2$ are non-universal parameters and we set $\xi_1=\xi_2 = 1$ for simplicity. Notably, the scaling function $\phi_F(x)$ only depends on the universality class. In \nbwo, the emergent CEP belongs to the 3D Ising universality class. Derived from the universal form Eq.~(\ref{Equ:FreeEnergy}), thermodynamics exhibits universal behaviors near the CEP. Particularly, the specific heat can be expressed as $C\equiv -T\frac{\partial^2 F}{\partial T^2} =t^{-\alpha}\phi_{C}(ht^{-\beta-\gamma})$. Considering the maximum condition $\frac{\partial C}{\partial T}=0$, the peak locations satisfy $ht^{-\beta-\gamma} = {\rm const.}$, implying $h\propto t^{\beta+\gamma}$, which is illustrated in Fig.~\ref{fig2}(c). Additionally, the magnetic susceptibility has a universal form $\chi\equiv -\frac{\partial^2 F}{\partial H^2} = t^{-\gamma}\phi_{\chi}(ht^{-\beta-\gamma})$. In the main text, the scaling function $\phi_{\chi}(x)$ is extracted from the data collapse of the measured susceptibility, in excellent agreement with the 3D Ising scaling function from Monte Carlo simulations, as shown in Fig.~\ref{fig2}(e). 

% ========================================
%                                   FIG E2
% ========================================
\begin{figure}[t!]
\centering
\includegraphics[width=0.75\linewidth]{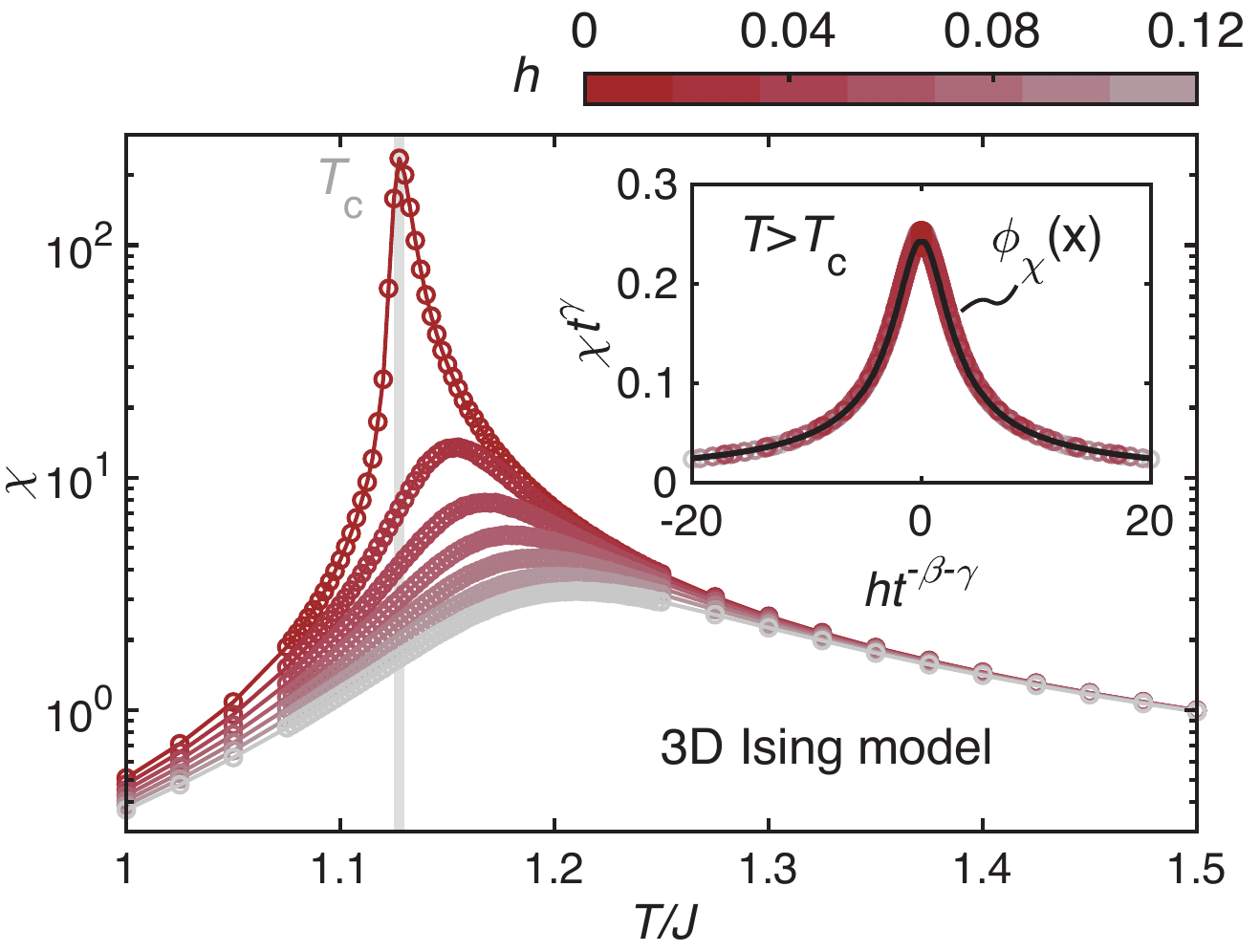}
\caption{The magnetic susceptibility $\chi$ of the 3D classical ferromagnetic Ising model for each fixed magnetic field $h$. Inset shows the supercritical data ($T>T_{\rm c}$) which can be collapsed onto the 3D-Ising scaling function $\phi_{\chi}(x)$ (black line). The Monte Carlo calculations are conducted on a cubic lattice of $N = 40^3$ Ising spins.}
\label{figE2}
\end{figure}

Below, we consider the Gr\"uneisen ratio $\Gamma_H \equiv \frac{1}{T}(\frac{\partial T}{\partial H})_S=-(\frac{\partial^2 F}{\partial H\partial T})/(T\frac{\partial^2 F}{\partial T^2})$, which characterizes the temperature variation during an adiabatic demagnetization process. Near the CEP, the Gr\"uneisen ratio also possesses a universal form 
\begin{equation}
    \Gamma_H = t^{1-\beta-\gamma}\phi_{\Gamma}(ht^{-\beta-\gamma}). 
    \label{UniGamma}
\end{equation} 
For an adiabatic demagnetization process, the peak/dip condition is $\left(\frac{\partial \Gamma_H}{\partial H}\right)_{S} = \frac{\partial \Gamma_H}{\partial H} + \left(\frac{\partial \Gamma_H}{\partial T}\right)\left(\frac{\partial T}{\partial H}\right)_S = \frac{\partial \Gamma_H}{\partial H} + \frac{\partial \Gamma_H}{\partial T}\Gamma_H T = 0$, using the definition $\Gamma_H\equiv \frac{1}{T}(\frac{\partial T}{\partial H})_S$. Considering the universal form Eq.~\ref{UniGamma}, the peak/dip condition can be expressed as $(1-\beta-\gamma)\phi_{\Gamma}^2(x) + \phi_{\Gamma}'(x)[\frac{1}{H_{\rm c}}-(\beta+\gamma)x\phi_{\Gamma}(x)] = 0$ with $x = ht^{-\beta-\gamma}$ and the corresponding zero point is $ht^{-\beta-\gamma}=x_0$. It means that the peak/dip values measured during the adiabatic demagnetization process satisfy a universal scaling $\Gamma_{H}^{\rm peak/dip} = t^{1-\beta-\gamma} \phi_{\Gamma}(x_0) \propto t^{1-\beta-\gamma}$ because $\phi_{\Gamma}(x_0)$ is a constant, as shown in Fig.~\ref{fig3}(d). 

\textit{Supercritical scaling function of 3D Ising universality class.---}
Since universal scaling functions characterize a universality class regardless of microscopic details, $\phi_{\chi}(x)$ for the 3D Ising class can be obtained by studying a canonical minimal model. We thus employ Monte Carlo simulations of the classical ferromagnetic (FM) Ising model on a cubic lattice, 
\begin{equation}
    H_{\rm Ising} = -J\sum_{<i,j>}S_i^z S_j^z - h\sum_i S_i^z, 
\end{equation}
where $J=1$ is the FM Ising coupling and $h$ denotes the longitudinal field. Our calculations are performed on a cubic lattice of size $N=40^3$. From the magnetic susceptibility $\chi$, we identify a critical temperature $T_{\rm c}\simeq 1.13$ for the finite-size system (see Fig.~\ref{figE2}). For $h\neq 0$, the phase transition evolves into a crossover characterized by a broad peak at higher temperatures. We collapse the $T>T_{\rm c}$ data and obtain the supercritical scaling function $\phi_{\chi}(x)$, which agrees excellently with the experimental results. Notably, non-universal parameters in Eq.~(\ref{Equ:FreeEnergy}) allow us to rescale the $x$ and $y$ axes to match the experimental results.

%==============================================
%                                       FIG E3
% ==============================================
\begin{figure}[t!]
\centering
\includegraphics[width=1\linewidth]{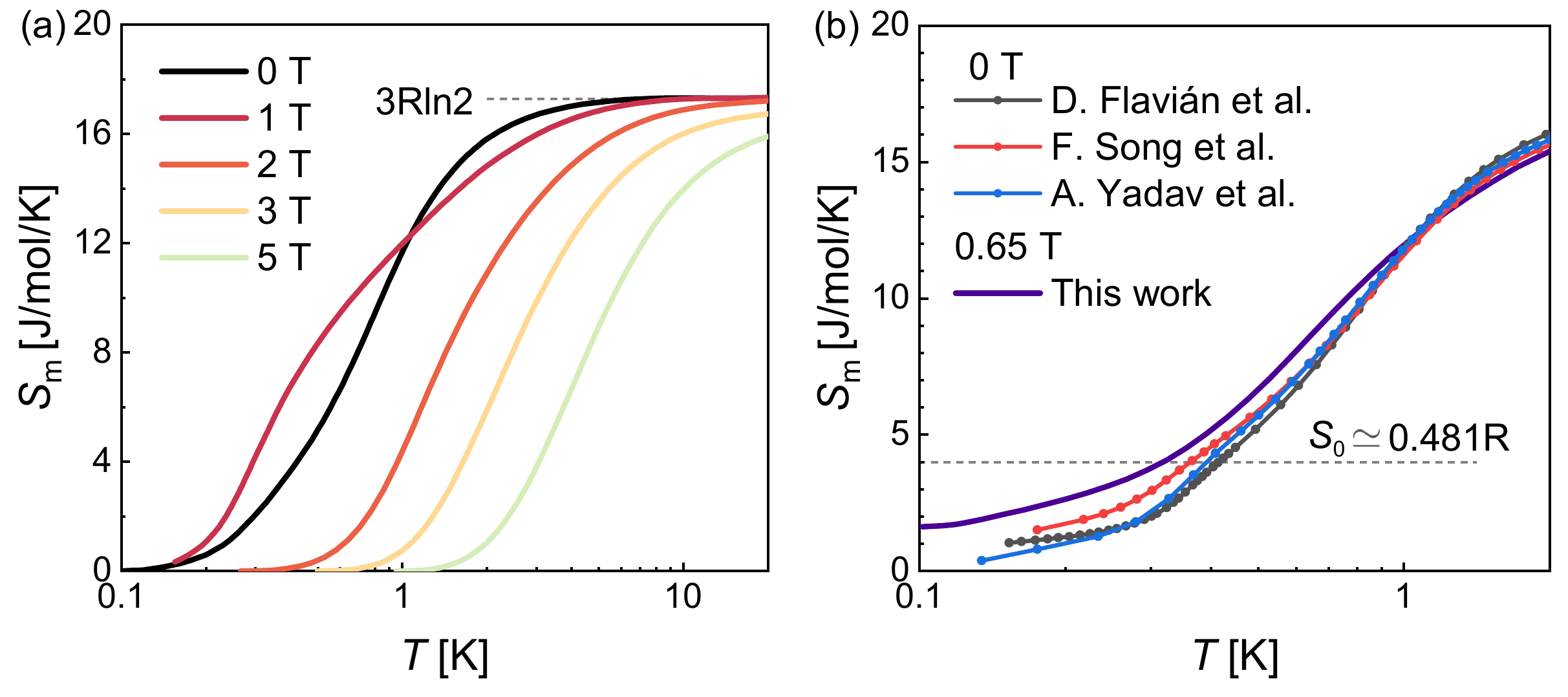}
\caption{
(a) Experimental entropy data obtained under different magnetic fields. (b) Entropy data from D. Flavi\'an \textit{et al.}~\cite{AZ2023PRB}, F. Song \textit{et al.}~\cite{Tian2023PRB}, and A. Yadav \textit{et al.}~\cite{Khuntia2025PRB} reveal a shoulder-like structure near $S_0 \simeq 0.481R$, attributed to topological domain-wall excitations. The low-temperature entropy increases significantly near $H_{\rm SF} \simeq 0.65$~T, where domain walls proliferate from even lower temperatures.
}
\label{figE3}
\end{figure}

% ================================================
%                                      FIG E4
% ================================================
\begin{figure}[t!]
\centering
\includegraphics[width=1\linewidth]{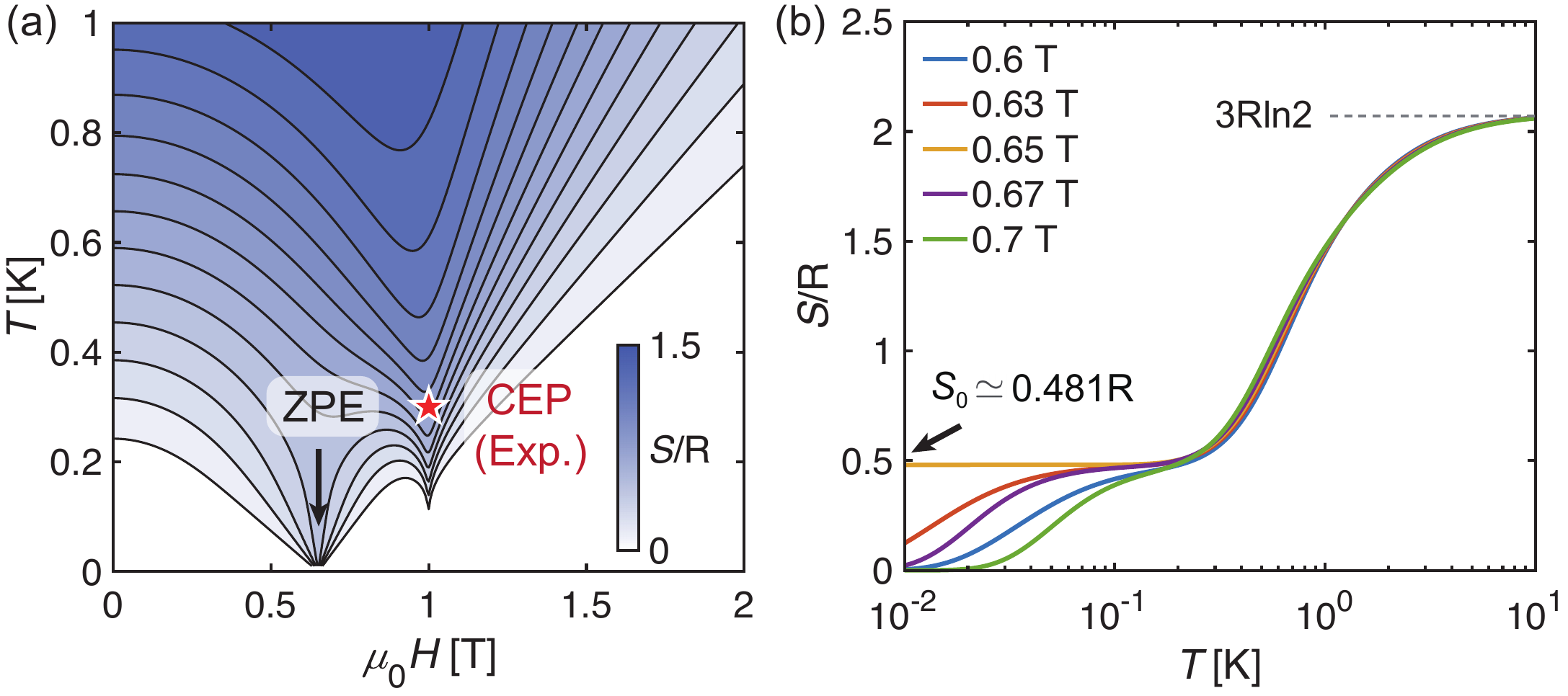}
\caption{
(a) Isentropic lines of the effective Ising tube model. The arrow illustrates the spin-flip transition field $\mu_0H_{\rm SF}\simeq 0.65$~T. The saturation field $\mu_0H_{\rm s}\simeq 1$~T in the Ising tube model is close to the critical field $\mu_0H_{\rm c}\simeq 1.04$~T in experiments. The red star marks the experimental CEP, which is absent in the 1D tube model. (b) Entropy curves near $H_{\rm SF}$, where the ZPE is $S_0\simeq 0.481R$. 
}
\label{figE4}
\end{figure}

\textit{{Experimental and simulated entropy results.---}}
Magnetic specific heat $C_{\rm m}$ is obtained by subtracting the nuclear and phonon contributions from the measured $C_p$~\cite{SM}. We then integrate $C_{\rm m}/T$ to obtain the magnetic entropy $S_{\rm m}$. Fig.~\ref{figE3}(a) shows the $S_{\rm m}(T)$ results under various fields. The zero-field magnetic entropy reaches a saturation value $3R \ln{2}$ at high temperatures. Figure~\ref{figE3}(b) displays a shoulder-like structure corresponding to a ZPE of $S_0 \simeq 0.481 R$. At the spin-flip field $\mu_0 H_{\rm SF} \simeq 0.65$~T, low-$T$ entropy is further enhanced due to the proliferation of topological domain-wall excitations.

Here, we compute the isentropes of the spiral Ising tube (SIT) model for \nbwo, using the transfer matrix method~\cite{AZ2025PRBL}, 
\begin{equation}
\begin{split}
    H_{\rm SIT} & = \sum_{i,j} (J_{\rm r} S_{i,j}^z S_{i+1,j}^z + J_{\rm b} S_{i,j}^z S_{i+1,(j-1){\rm mod}3}^z) \\
    & - g_{zz}\mu_0 \mu_{\rm B} \mathbf{H \cdot \hat{z}} \sum_{i,j} S_{i,j}^z,
\end{split}
\end{equation}
where $i$ labels the layer along the $c$ axis and $j=\{1,2,3\}$ labels three spins in a triangle in the $ab$ plane. As shown in Fig.~\ref{fig1}(c), the red and blue bonds represent two different Ising couplings, namely, $J_{\rm r} \simeq -0.084$~meV and $J_{\rm b} \simeq 0.24$~meV, respectively~\cite{AZ2025PRBL}. The angle between the magnetic field and Ising $z$ axis is 54$^{\circ}$ and we set the $g_{zz}\simeq 7.06$. 

The simulated isentropes are shown in Fig.~\ref{figE4}(a), from which we find two prominent dips, one at the saturation field $\mu_0 H_{\rm s}\simeq 1$~T and the other at the spin-flip field $\mu_0 H_{\rm SF} \simeq 0.65$~T. Although the SIT model accounts for the low-temperature cooling via ZPE [$S_0 \simeq 0.481 R$, Fig.~\ref{figE4}(b)] near $H_{\rm SF}$, the 1D model cannot capture the singular behaviors near the CEP ($\mu_0H_{\rm c} \simeq 1.04$ T, $T_{\rm c} \simeq 0.3$ K) observed in experiments. These supercritical scaling behaviors are inherently 3D, rather than 1D, arising from both intra- and inter-tube couplings~\cite{SM}. 

In the spiral AF phase, there exists a three-rung period along the spin tube, with the ground-state energy per site $\varepsilon_{\rm AF} = S^2(-J_{\rm b} - |J_{\rm r}|/3)$ and $S=1/2$. The three spins within the same layer on the $a$-$b$ plane can be collectively described by a single effective spin, which represents either the (up-up-down) UUD or (down-down-up) DDU configuration. The effective spins form an AF chain~\cite{SM}. At the spin-flip field $H_{\rm SF} = 2S(J_{\rm b} - |J_{\rm r}|)$, domain walls, as topological defects, emerge and can move freely without energy cost, i.e., $\varepsilon_{\rm DW} = 2S^2(J_{\rm b} - |J_{\rm r}|) - H_{\rm SF}S = 0$. For an $N$-site effective spin chain, the degeneracy is $\Omega = \sum_{i = N/2}^N \frac{(i+1)!}{(N-i)!(2i-N+1)!}$. Consequently, the ZPE per unit cell at $H_{\rm SF}$ reads $S_0 = \lim_{N\to\infty} \frac{1}{N} \ln {\Omega} \simeq 0.481$. For $H \gtrsim H_{\rm SF}$, there exists a domain wall on every bond, and the system is in the spiral plateau phase (all UUD) of the original Ising tube.

\end{document}